\documentclass[a4paper,11pt]{article}
\usepackage{textgreek}
\usepackage[utf8]{inputenc}
\usepackage{graphicx}
\usepackage{amssymb,amsmath,amsthm}
\usepackage{subfigure,subfloat}
\usepackage{anysize}
\marginsize{4cm}{4cm}{5cm}{5cm}
\usepackage{tikz}
\usepackage{authblk}
\usetikzlibrary{calc,patterns,angles,quotes} 
\usepackage{url}
\usepackage{color}

\graphicspath{{figs/}}

\title{Fungal architecture}
\author[1]{Andrew Adamatzky}
\affil[1]{Unconventional Computing Laboratory, Department of Computer Science, University of the West of England, Bristol, UK}
\author[2]{Phil Ayres}
\affil[2]{The Centre for Information Technology and Architecture, The Royal Danish Academy of Fine Arts
Schools of Architecture, Design and Conservation, Copenhagen, Denmark}
\author[3]{Gianluca Belotti}
\affil[3]{Mogu S.r.l., Inarzo, Italy}
\author[4]{Han W\"{o}sten}
\affil[4]{Microbiology, Department of Biology, Utrecht University, Utrecht, The Netherlands}
\date{}

\date{}

\begin{document}

\maketitle

\begin{abstract}
\noindent

\vspace{2mm}
\noindent
As one of the primary consumers of environmental resource, the building industry faces unprecedented challenges in
needing to reduce the environmental impact of current consumption practices. This applies to both the construction of the
built environment and resource consumption during its occupation and use. Where incremental improvements to current
practices can be realised, the net benefits are often far outstripped by the burgeoning demands of rapidly increasing
population growth and urbanisation. Against the backdrop of this grand societal challenge, it is necessary to explore
approaches that envision a paradigm shift in how material is sourced, processed and assembled to address the magnitude
of these challenges in a truly sustainable way, and which can even provide added value. We propose to develop a structural
substrate by using live fungal mycelium, functionalise the substrate with nanoparticles and polymers to make a
mycelium-based electronics, implement sensorial fusion and decision making in the fungal electronics and to
growing monolithic buildings from the functionalised fungal substrate. Fungal buildings will self-grow, build, and repair
themselves subject to substrate supplied, use natural adaptation to the environment, sense all what human can sense. 

\vspace{3mm}

\end{abstract}

\section{Introduction}

As one of the primary consumers of environmental resource, the building industry faces unprecedented challenges to reduce the environmental impact of current consumption practices. This applies to the construction of the built environment and resource consumption during its occupation and use. Where incremental improvements to current practices can be realised, the net benefits are often far outstripped by the demands of rapidly increasing population growth and urbanisation. Against the backdrop of this grand societal challenge, it is necessary to explore approaches that envision a paradigm shift in how material is sourced, processed, and assembled to address the magnitude of these challenges in a sustainable way, with added value. We propose to develop a structural substrate by using live fungal mycelium, functionalise the substrate with nanoparticles and polymers to make a mycelium-based electronics, implement sensorial fusion and decision making in the fungal electronics and to grow monolithic buildings from the functionalised fungal composites. Fungal buildings will self-grow to target geometries, self repair, remediate waste products, sense and naturally adapt to the environment.  The buildings will have low production and running costs, will not require a substantial workforce to build, they are ecologically friendly and waste can be returned to nature when no longer used. This building cycle will radically decrease or even nullify the environmental costs of building material production.  Fungal architecture presents a radical vision of architecture that can rapidly, for living beings, grow in-situ, providing 'free’ material sourcing: unprecedented possibilities in terms of technical, aesthetical and sustainable solutions. Mycelium networks will be computationally active, giving rise to entirely new biologically founded functionalities for architectural artefacts and materials, such as self-regulation, adaptation, decision making, autonomous growth, searching and self-repair --- adding new advantages and value to architectural artefacts and the environment, and providing a radically alternative paradigm to the state-of-the-art in ‘intelligent buildings’ which are heavily reliant upon technical infrastructures. 

\section{State of the Art}

\begin{figure}[!h]
    \centering
   \includegraphics[width=0.99\textwidth]{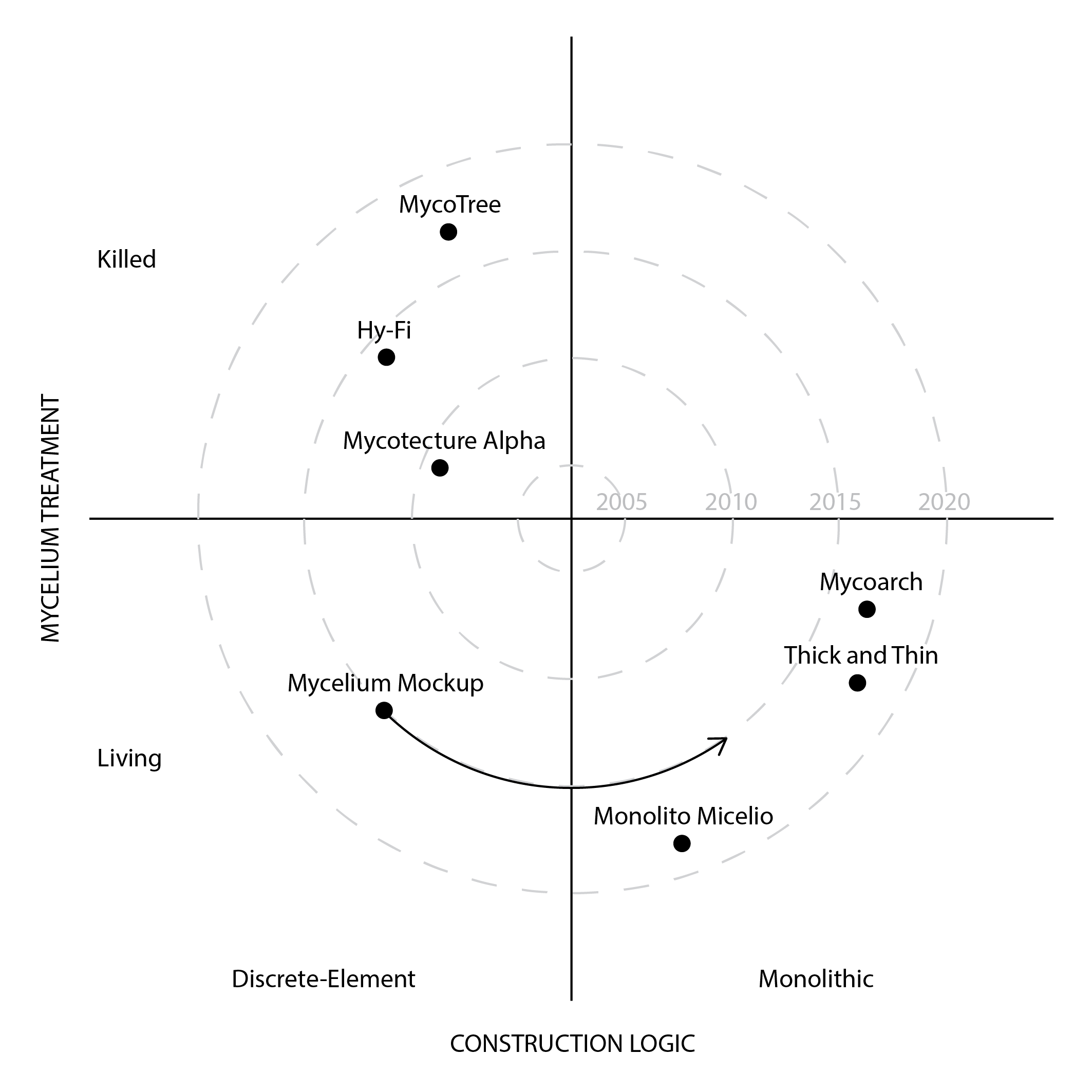}
    \caption{Taxonomy of architectural projects employing mycelium composites in structural applications}
    \label{fig:taxonomyofprojects}
\end{figure}

Research in the production and application of mycelium composites within, or allied to, the construction industry is a nascent field that remains under investigated. Here, we survey prototypes and projects that explicitly consider structural applications for mycelium composites in building construction. We find that there are very few examples reported in the literature, nevertheless, a distinction is drawn through production approaches that seek to grow monolithic structures and those based on discrete-element assemblies. A second distinction is drawn between production approaches that kill the mycelium during the production process and those that seek to maintain the mycelium as a living organism. These two distinctions provide the basis of a taxonomy for classifying the projects, as shown in Fig.~\ref{fig:taxonomyofprojects}  

In 2009, Philip Ross exhibited `Mycotecture Alpha', a composition of discrete mycelium composite blocks arranged in a barrel vault configuration reminiscent of stone construction~\cite{ross2016your}.
The `Hy-Fi' installation, exhibited at MOMA PS1, New York, in 2014, was a 13~m tower composed of hand-stacked pre-fabricated and heat treated mycelium bricks in combination with an ancillary structure of steel and timber~\cite{saporta2015design}. This temporary installation was externally located, thus subject to, and surviving, the vagaries of environmental loads over the course of approximately two months between July and September. 
Exhibited at the 2017 Seoul Biennale, the `MycoTree' installation demonstrates greater complexity in the moulding approach~\cite{heisel2017design, heisel2018design}. The design intent was to produce a compression-only branching structure. This was realised with a family of different element types, including branching nodes with a tetrahedral topology and triangular cross-section.

Common to all of the projects cited above is the production of the discrete element by growing a mycelium matrix to interlock with a substrate, resulting in a composite bulk material. Inoculated substrate is geometrically contained in moulds over the growth period. When the mycelium has sufficiently colonised the substrate, the element is de-moulded and heat-treated to remove residual moisture and kill the mycelium.

In contrast to these pre-fabricated discrete-element approaches, Dessi-Olive examines the potentials of in-situ monolithic growth through a series of 1:1 artefacts~\cite{DessiOlive}. Central to this approach is the use of ‘lost’ internal reinforcing systems in combination with external removable formworks – making the production and assembly methodology analogous to that of in-situ concrete casting, rather than that of masonry as in the case of the previously cited works. This methodology is developed through three prototypes – ‘Mycoarch’ (2017), ‘Thick and Thin’ (2018), ‘Monolito Micelio’ (2018). In all cases, the mycelium composite was air-dried rather than heat-treated.

Dahmen reports on a hybrid approach employed by AFJD Studio to produce a temporary wall installation called ‘Mycelium Mockup’(2015)~\cite{dahmen2016soft}. Discrete brick elements were produced and de-moulded without a growth arresting treatment. These living blocks were arranged in the target formation and continued to grow, establishing cross-links with neighbouring bricks and transforming a discrete-element assembly into a monolithic assembly.

In summary, common to all approaches is the privileging of compression based structural geometries due to the mechanical properties of the mycelium composite. Monolithic construction approaches present challenges in achieving consistency of controlled production, which may have an impact on material performance, particularly as scale increases. In contrast, discrete-element approaches offer greater control in production but are reliant upon subsequent assembly methods. Heat treatment of mycelium composites to kill the mycelium provide a means of producing inert material, however, living material has properties that could offer new active properties to architectural constructions such as self-healing, self-repair, partial self-assembly and decision making – all of which remain under-investigated to date.

\section{Ideas}

Our overarching goal of designing and bio-manufacturing a sensing and computing building with fungi will be achieved via the specific objectives described below: 
\begin{itemize}
    \item  Biofabrication: Cultivation of large (metre length scale) living mycelium.  This will result in developing new growth and cultivation protocols for large scale living mycelium networks and a success can be measured by the mycelium has fully colonised the targeted structure and remains living and healthy. Biofabrication can be defined as the production of complex living and non-living biological products from raw materials such as living cells or molecules. In the long term, biofabrication can dramatically transform traditional industries becoming a new paradigm for 21st century manufacturing.
    \item Functionalizing: Changing electrical and mechanical properties of mycelium network. This objective will lead us towards development of methodologies for localising specific synthetic properties within mycelium, where functionalized mycelium is capable for conductance and processing of information.
\item Computing: Implementation of information processing on mycelial networks. New methodologies and experimental implementations of hybrid living electronic and computing circuits will applied in the implementation of interfacing, sensorial fusion and logical operations.
\item Designing: Development of design rules and construction logics. Disruptive architectural tools/methods for designing and constructing sustainable architecture using processes of natural growth functionalised to attain computational capabilities will lead to prototypes of fungal architectures which actively respond to environmental changes and stimulation from inhabitants.
\end{itemize}

\section{Methodological considerations}

What species of fungi to use?  We would suggest that candidates would be representatives of \emph{P. citrinopileatus}, \emph{P. ostreatus}, \emph{L. edodes}, \emph{A. bisporus}, \emph{F. velutipes}, \emph{M. oreades}, \emph{A. rubescens}, \emph{C. geotropa}, \emph{G. applanatum}, \emph{G. lucidum}, and \emph{S. commune}. Based on the mechanical properties of mycelium composites, and metal tolerance studies and sorption capacity, two or three species could be selected for final prototypes of fungal architectures.  The electronic functionality, e.g. establishing conductive pathways and non-linear electrical circuits could be provided with poly(3,4-ethylenedioxythiophene) and polystyrene sulfonate (PEDOT-PSS) will be added to the substrate and bind/sorbed by the mycelium. Additionally, graphene/PEDOT:PSS dispersion, Plexcore OC AQ-1250, polyaniline, polypyrrole.
There is a substantial body of work already one producing unconventional computing devices with slime mould \emph{P. polycephlaum}~\cite{cifarelli2014non,romeo2015bio,dimonte2016physarum,de2015conducting,battistoni2018organic} which supports our ideas that fungal micelium can be also functionalised. 
Nanomaterials, carbon nanotubes, graphene oxide, aluminium oxide, calcium phosphate, that affect electrical properties of mycelium will be added to the substrate to program functional networks. 

Would functionalisation with metal be toxic for mushrooms? Fungi differ in metal/polymer tolerance. \emph{Pleurotus} species show higher resistance to the heavy metals, copper, cadmium, zinc, nickel, cobalt, and mercury than other mushroom species~\cite{sanglimsuwan1993resistance,baldrian2010effect} while e.g. \emph{F. velupites} and \emph{A. bisporus} are rather sensitive~\cite{kulshreshtha2014mushroom}. 

To analyse metal binding or sorption. Fungi can  will be grown in the presence of 13C-labelled glucose, if needed also D-xylose, and L-arabinose are suitable. Cell walls will be isolated/extracted with water, alkali and acid, respectively. Solid-state NMR spectra of each of the fractions will be subjected to dipolar- and scalar-based multidimensional correlation under Magic Angle Spinning conditions. Conductive molecule binding to each of the fractions will be correlated to the composition of each of the extracts. This will indicate which molecules are the main players in binding of these compounds. This will be confirmed using purified polysaccharides or strains with overexpressed cell wall components.

 To program topology of mycelium network, directionality of growth and degrees of branching at any given site we will be modifying varying nutritional conditions and temperature~\cite{boddy1999fractal, hoa2015effects, rayner1991challenge,regalado1996origins}. Thus, e.g. a  degree of branching is proportional to concentration of nutrients~\cite{ritz1995growth}, a range of chemical and physical stimuli~\cite{bahn2007sensing}.   Chemical control/programming of mycelial architecture and behaviour will be implemented via application of attractants/repellents in the substrate and/or in the air. Chemoattractants are oxidised lipid, carbohydrates, peptones, and selected amino-acids. Chemo-repellents are sucrose, tryptophan and salt. We should also keep in mind a possibility of stimulation with volatiles and light.  Based on our previous experience with slime mould~\cite{delacycostello2013assessing,adamatzky2012physarum} the compounds will be selected to represent four chemical classes --- aromatic, monoterpenes, sesquiterpenes and fatty acid derivatives established in the evolution of floral and insect volatiles. 

 Another way to control the geometrical structure of mycelium would be to  geometrically constrain it~\cite{hanson2006fungi, held2008examining, held2009fungal, held2010microfluidics,held2011probing} by applying pre-designed templates. 
 
 Hardwiring the mycelium network with electricity is amongst most promising tools because it does require diffusion of chemicals (slow process) or templating (mechanical alteration).  A feasibility of such programming similar living networks has been demonstrated in~\cite{whiting2016towards} with slime \emph{P. polycephalum}: high amplitude high frequency  voltage applied between two electrodes in a network of protoplasmic tubes of \emph{P. polycephalum} led to abandonment of the stimulated protoplasmic without affecting the non stimulated tubes and low amplitude low frequency voltage applied between two electrodes in the network enhance the stimulated tube and encourages abandonment of other tubes~\cite{whiting2016towards}. In experiments described in~\cite{takaki2009effect} a crop of \emph{P. nameko}  was increased by short-pulse stimulation of the substrate with high voltage (120~kV with 100~ns pulse width). This indicates that a very short high-energy electrical impulses might lead to increased branching of the mycelium.

 Electrical properties of mycelial networks will be explored in terms of resistance (including thermistors and memristors), capacitance, transistor like functionality (control of voltages or currents so as to accomplish gain or switching action in a circuit), operational amplification, multiplication, fixed-function generation. Implementations to consider would include active elements analog computers, without amplification and passive computers, where amplification of signal is necessary.
 
Architectural design is amongst strong keys  of the fungal architectures. Our innovative structural and shaping methodology uses Kagome (triaxial) woven structures as a combined stay-in-place mold and reinforcement for the mycelium/substrate composite~\cite{ayres2018beyond}.

The Kagome structures are designed using computational methods that combine environmental and structural simulation to inform the topology and geometry of design meshes. Using the Medial construction method, these meshes are converted into Kagome patterns and relaxed using physics-based simulation methods to ensure the design is producible using straight strips of material. Throughout the project we will employ an iterative approach to physical prototyping. Initial prototypes will focus on gaining design feedback and achieving success at component scale before attempting larger structures. 

\section{Illustrative study}

\begin{figure}[!tbp]
    \centering
    \subfigure[]{\includegraphics[width=0.24\textwidth]{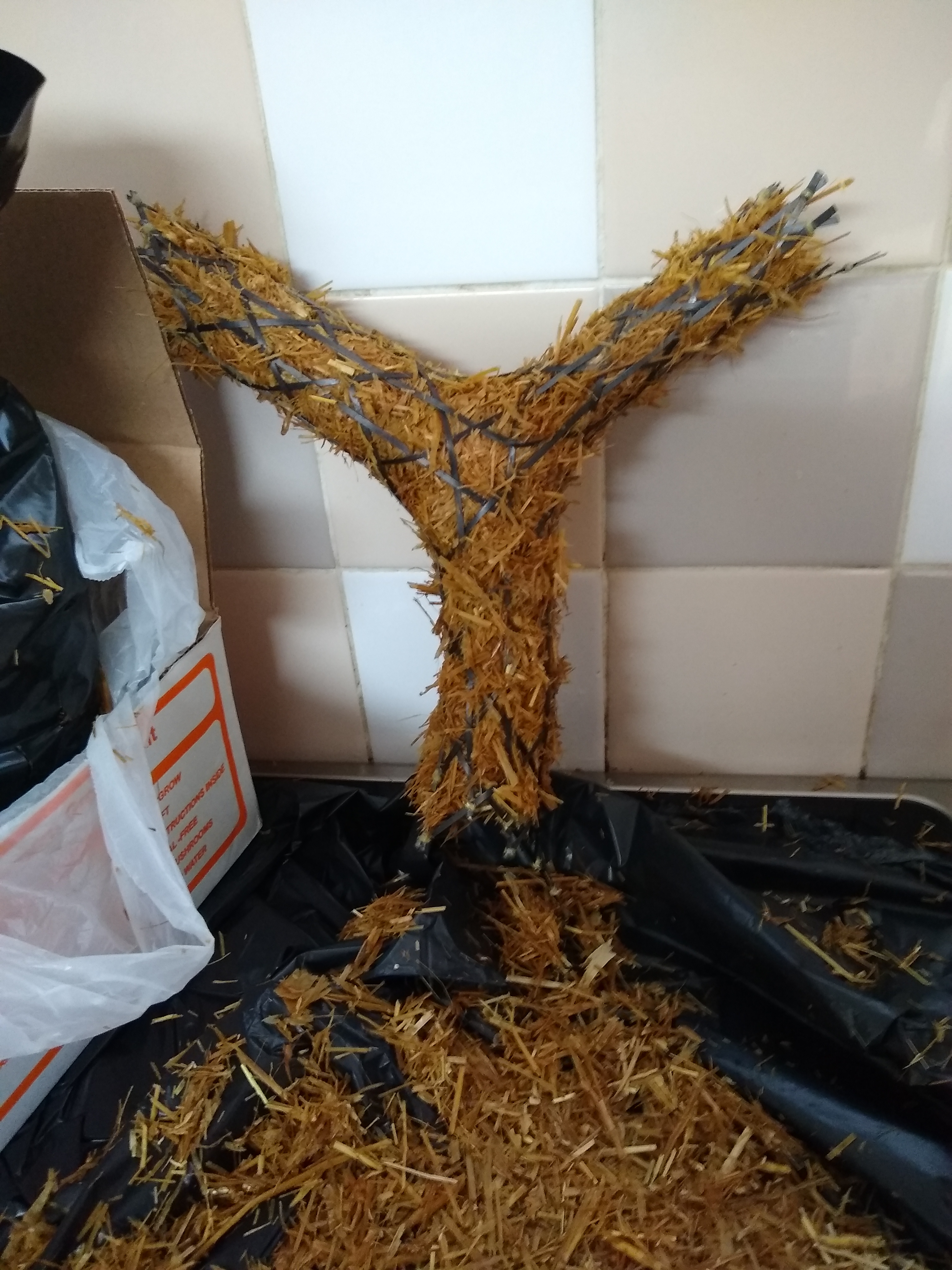}\label{fig:exoskeletonstuffed}}
    \subfigure[]{\includegraphics[width=0.24\textwidth]{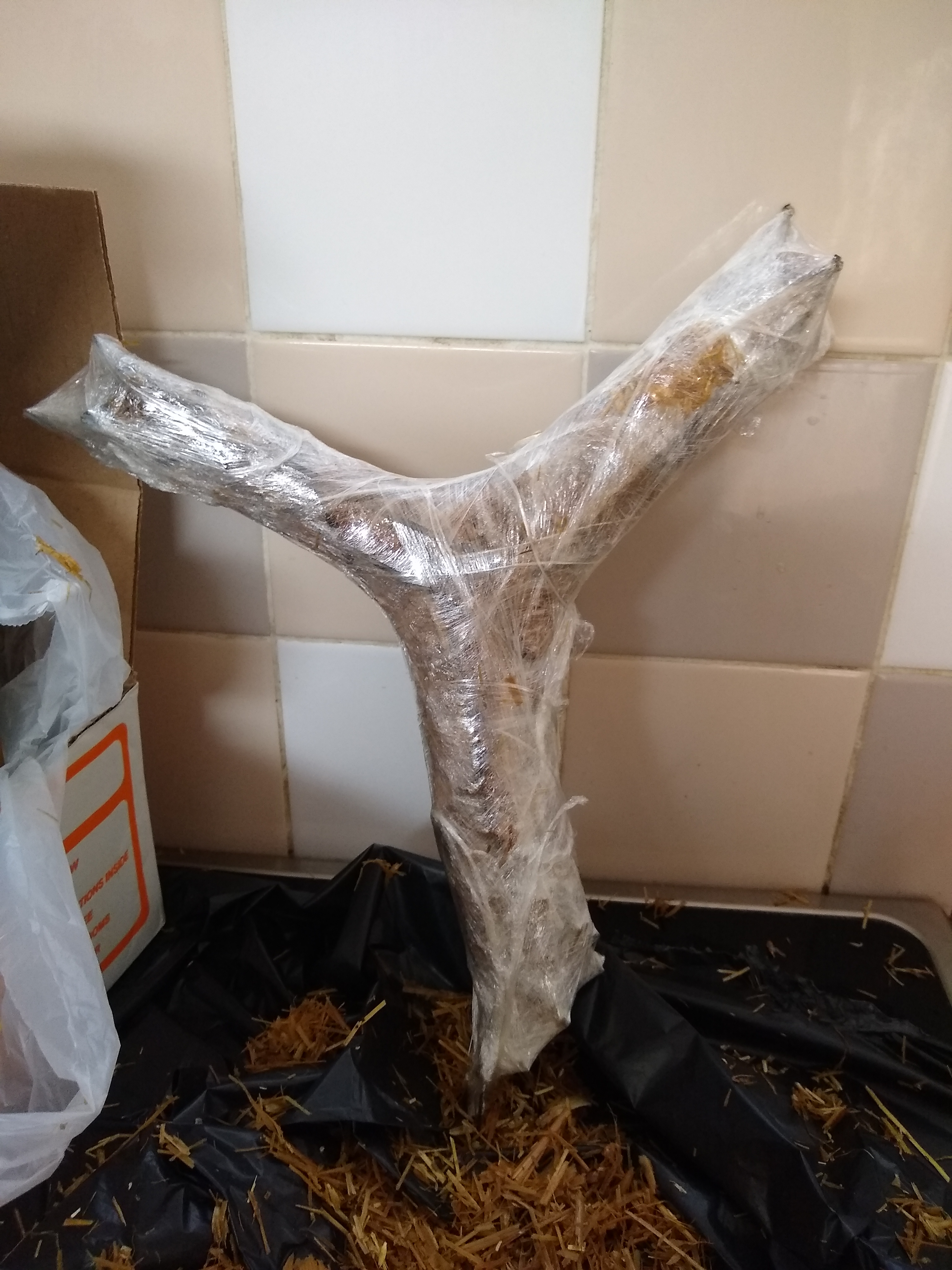}\label{fig:exoskeletonwrapped}}
    %\subfigure[]{\includegraphics[width=0.24\textwidth]{IMG_20181227_130454586}}
    \subfigure[]{\includegraphics[width=0.40\textwidth]{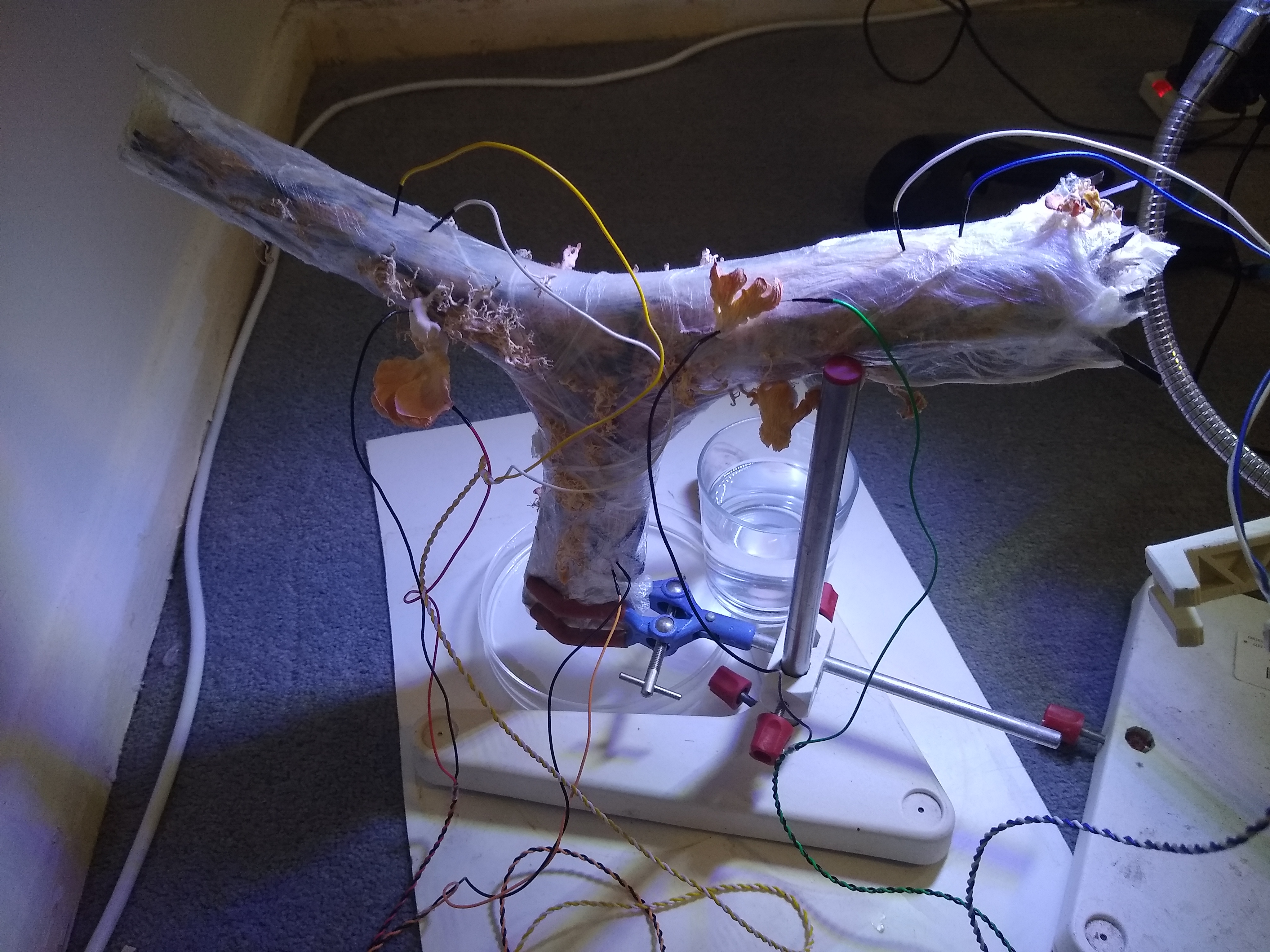}\label{fig:electrodesinsert}}
     \subfigure[]{\includegraphics[width=0.32\textwidth]{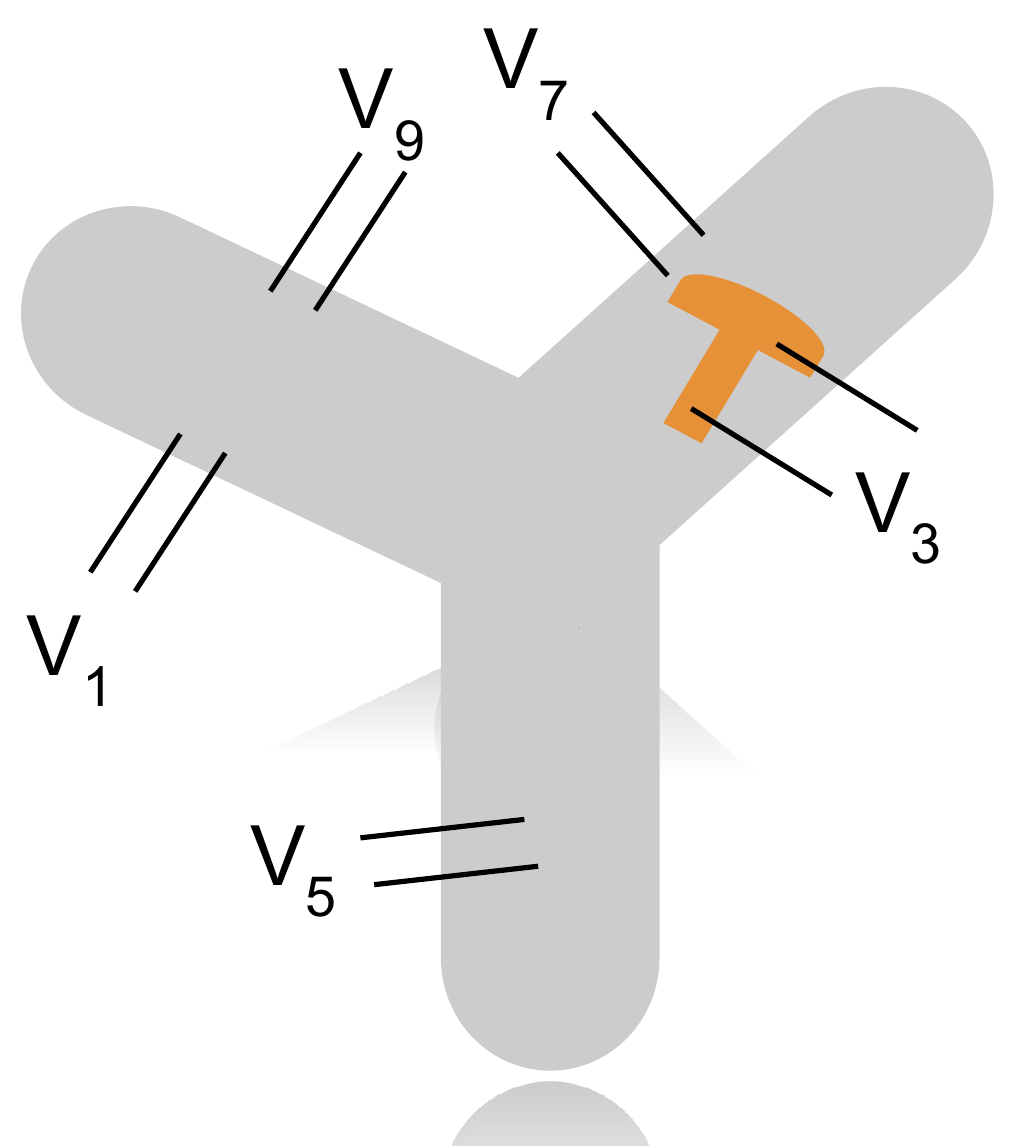}\label{fig:schemeofelectrodes}}
    \subfigure[]{\includegraphics[width=0.32\textwidth]{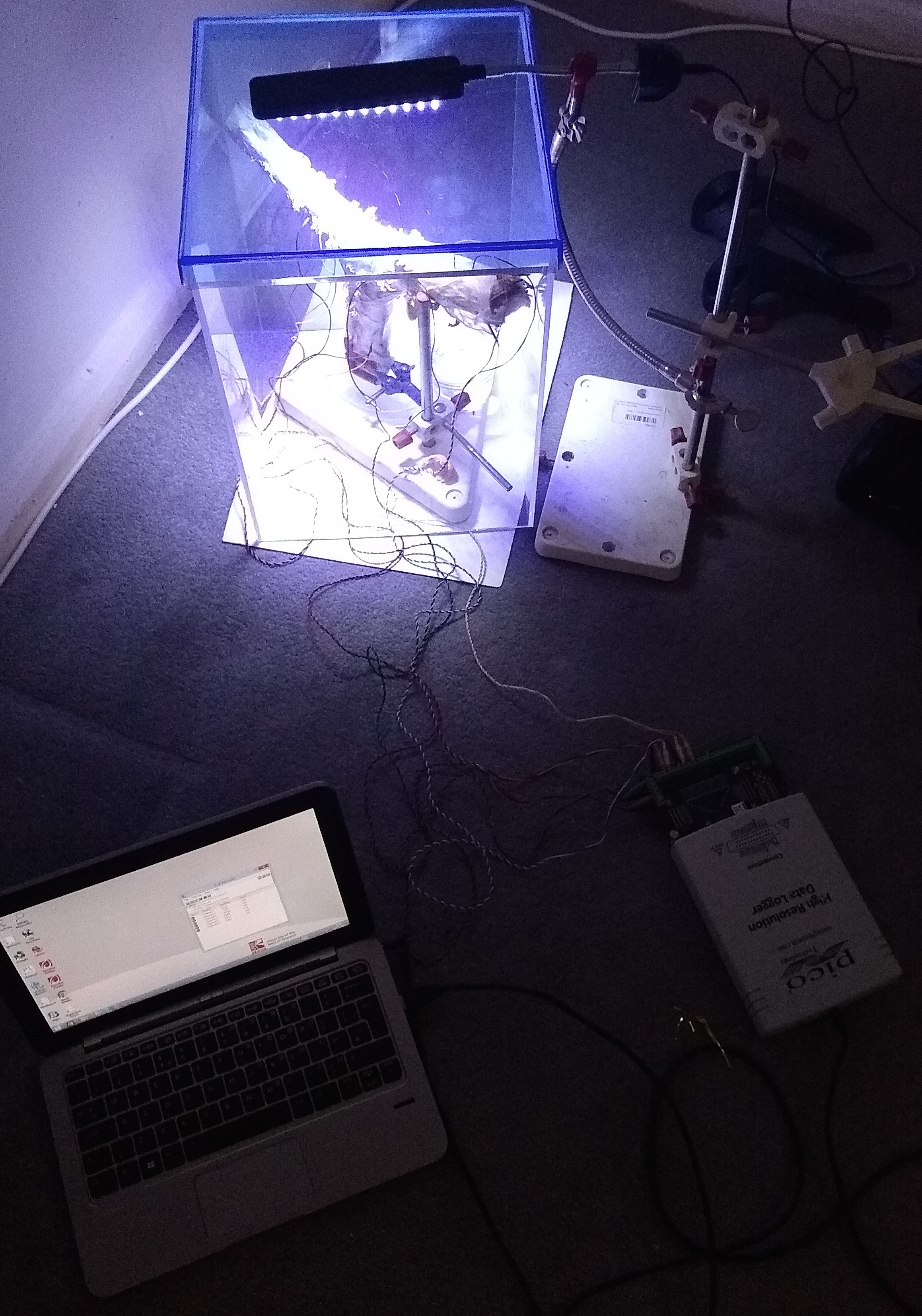}\label{fig:overallsetup}}
    \caption{Experimental setup. (a)~exoskeleton is stuffed with a growth substrate, hay, and (b) wrapped into a cling film, to keep humidity constant, holes are punched along the stem and branches of the structure. When first fruits appeared, we undertook recording of electrical activity by inserting electrodes in fruit body and substrate, as illustrated in (c) and (d); photo of the overall setup is shown in (e). $V_1$, \ldots, $V_9$ are differences of electrical potential recorded between pairs of electrodes.}
    \label{fig:experimental}
\end{figure}

\begin{figure}[!tbp]
    \centering
   \includegraphics[width=0.99\textwidth]{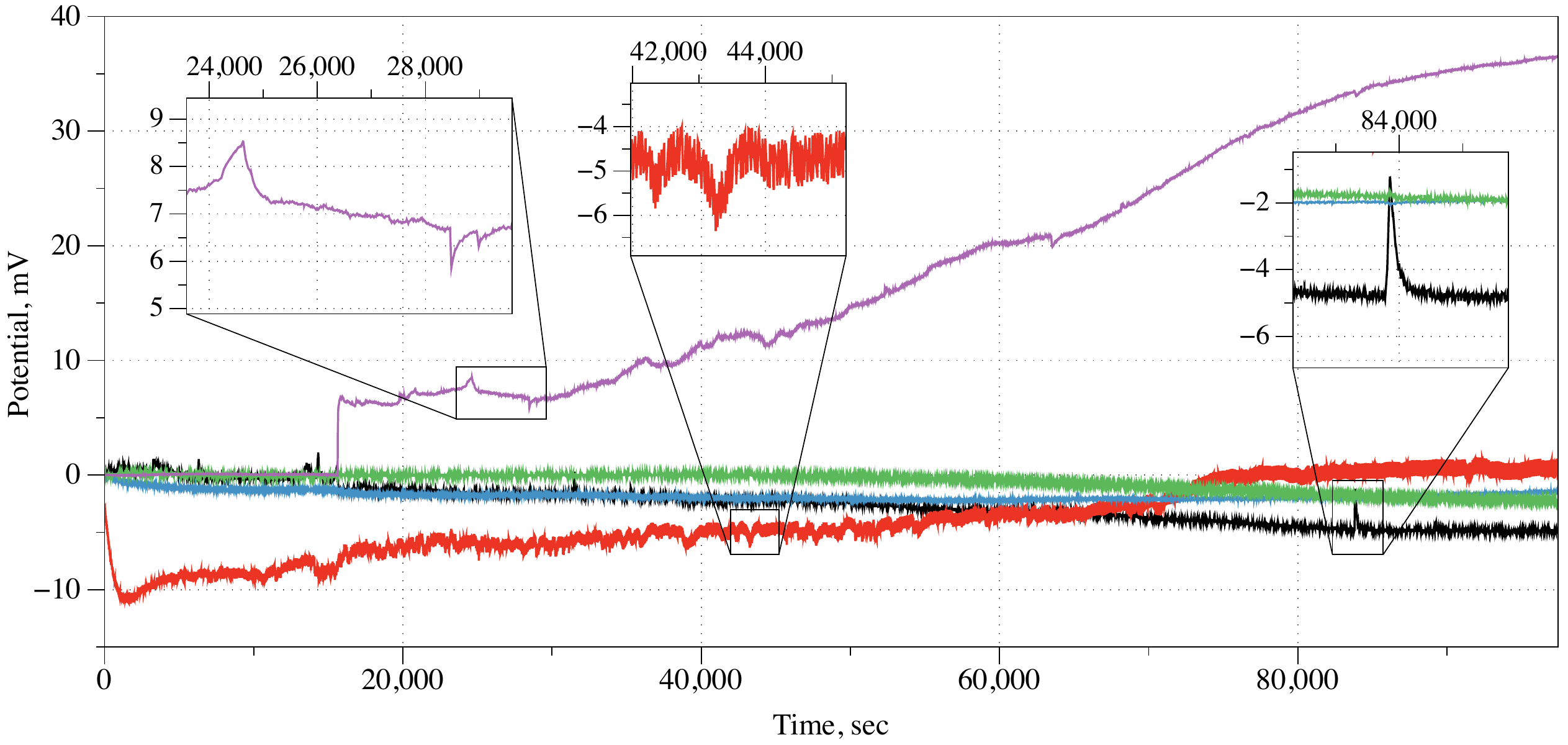}
    \caption{Electrical potential recorded during 25 hours.  See positions of electrode pairs in Fig.~\ref{fig:schemeofelectrodes}. 
    Electrical potential in  $V_1$ is black, $V_3$ is red, $V_5$ is cyan, $V_7$ green, $V_9$ magenta.}
    \label{fig:recordings1}
\end{figure}

The overall idea is illustrated, at a smaller scale in Figs.~\ref{fig:experimental} and \ref{fig:recordings1}. The weave pattern skeleton~\cite{ayres2018beyond} is filled with a grow substrate, mixed with mycelium of \emph{P. djamor} (Fig.~\ref{fig:exoskeletonstuffed}) and wrapped in cellophane film with holes punched in it (Fig.~\ref{fig:exoskeletonwrapped}).  Pairs of electrodes  (sub-dermal needle electrodes with twisted cable, Spes Medica Italy) have been inserted either in fruiting bodies of directly into the substrate (Fig.~\ref{fig:electrodesinsert}). In each pair one electrode was a reference electrode and another recording electrodes. Electrical activity of fruit bodies was recorded with ADC-24
High Resolution Data Logger\footnote{Pico Technology, St Neots, Cambridgeshire, UK}. The data logger  employs differential inputs, galvanic isolation and software-selectable sample rates --- these contribute to a superior noise-free 
resolution; its 24-bit A/D converter maintains a gain error of 0.1\%. Its 
input impedance is 2~M$\Omega$ for differential inputs, and offset error is 36$\mu$V 
in $\pm$ 1250~mV range use. We recorded the electrical activity one sample per second; during the recording the logger made as many measurements as possible (typically up 600) per second then saved average value. Positions of exact electrodes are shown on the scheme in  Fig.~\ref{fig:schemeofelectrodes} and overall recording setup in Fig.~\ref{fig:overallsetup}. 

\begin{figure}[!tbp]
    \centering
       \subfigure[]{\includegraphics[width=0.82\textwidth]{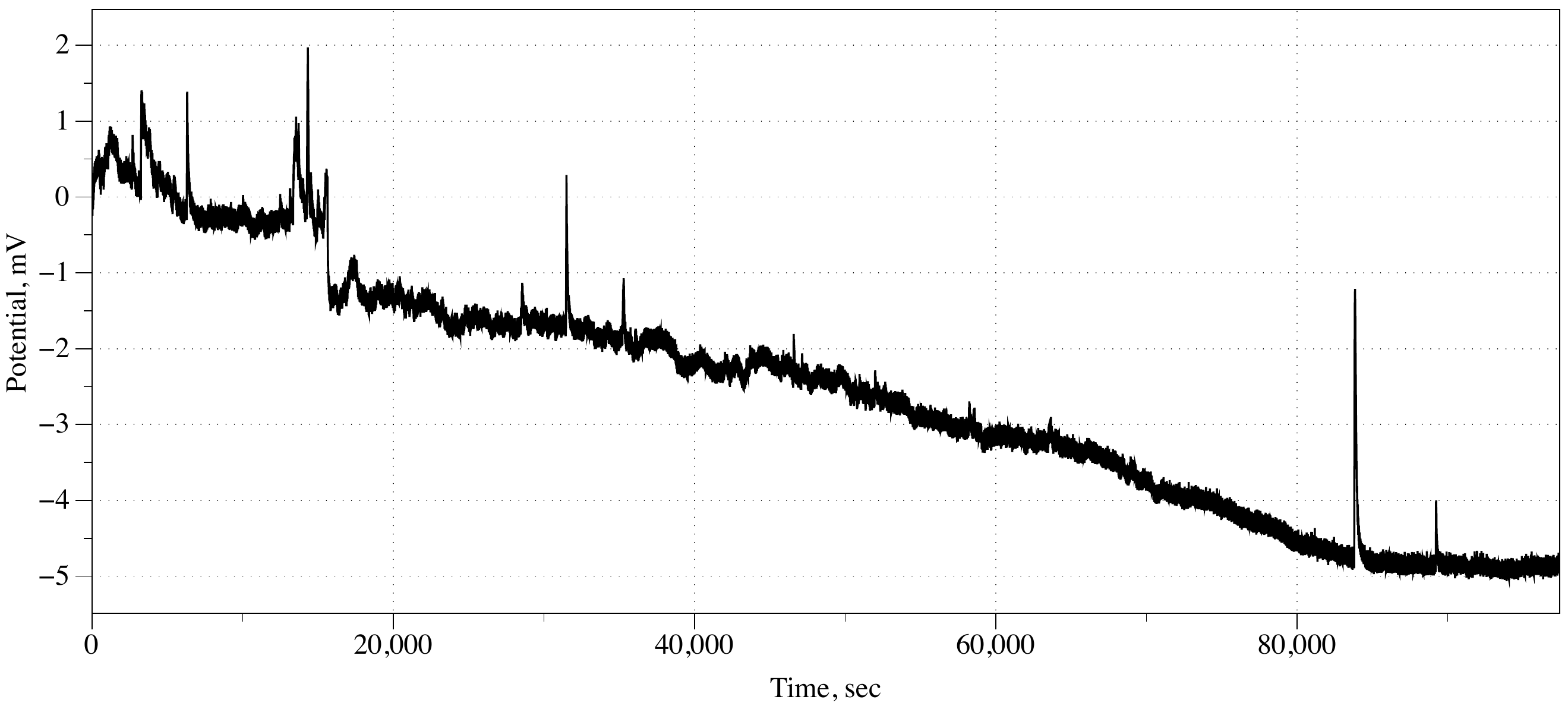}\label{V1}}
       \subfigure[]{\includegraphics[width=0.82\textwidth]{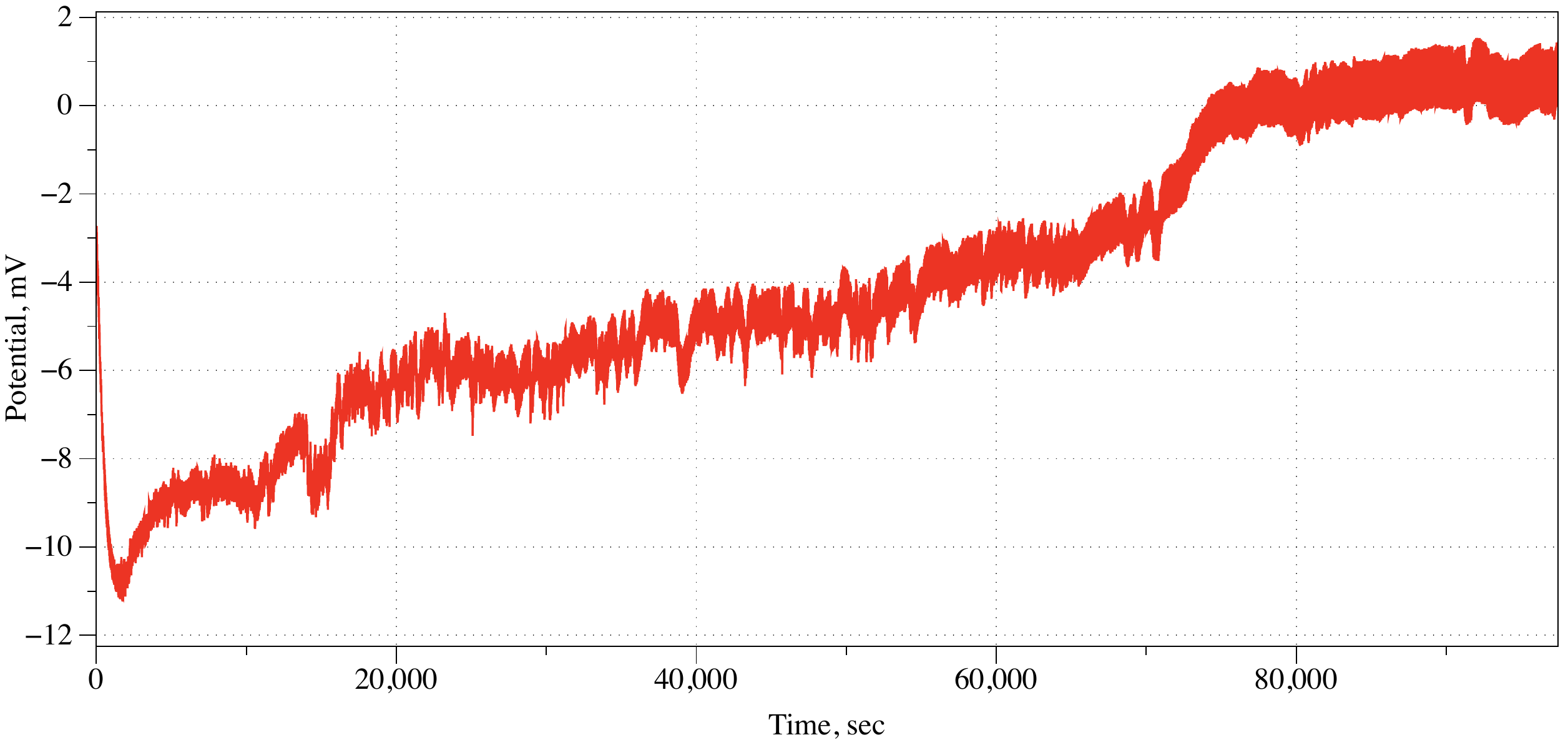}\label{V3}}
       \subfigure[]{\includegraphics[width=0.82\textwidth]{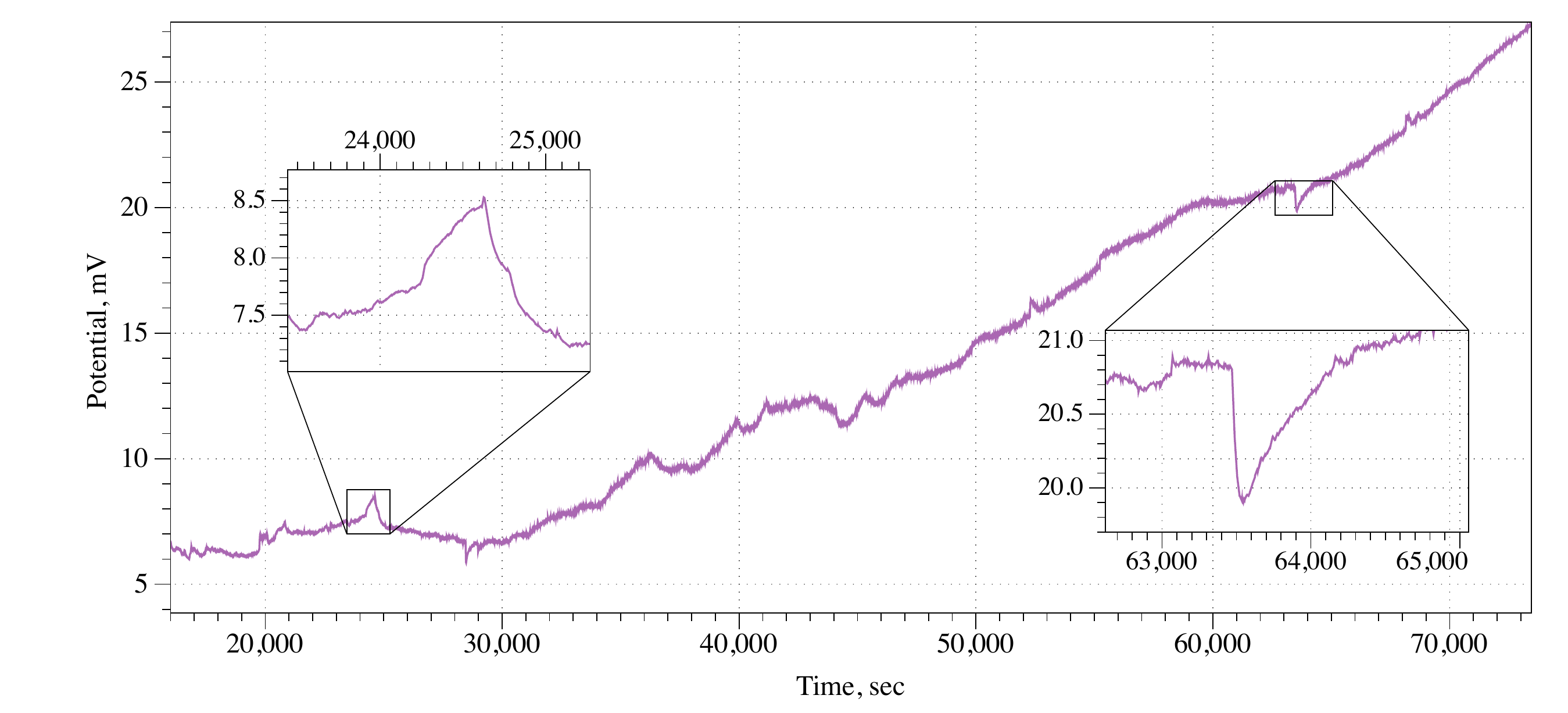}\label{V9}}
    \caption{Zoomed recordings of electrical potential on electrodes pairs (a)~$V_1$, (b)~$V_3$, and (c)~$V_9$.}
    \label{fig:recordings2}
\end{figure}

Electrical potential record show a substantial variability as evidenced in Fig.~\ref{fig:recordings1}. Let us consider in details, see fragments of the recordings zoomed in Fig.\ref{fig:recordings2}. The potential on $V_1$ (Fig.~\ref{V1}) drops by 5~mV during period of nearly 25 hours. We observe six spikes with amplitude over $1~mV$, over this sample average amplitude is 1.97~mV ($sigma=0.9$). The longest spike 410~min; other spikes ranged from c. 10~min sec to 30~min, with average 17~min ($\sigma=6.5$~min). The electrodes pair $V_3$ (Fig.~\ref{V3}), despite being a noisy channel, exhibits several well identifiable spikes. The average amplitude of these spikes  is 1.4~mV ($\sigma=0.33$~mV) and average period $37$~min ($\sigma=17$~min). Recordings of $V_9$ demonstrate few spikes with amplitude higher than 1~mV (Fig.~\ref{V9}).  The three spikes observed on $V_9$ have an average amplitude of 1.08~mV ($\sigma=0.072$~mV) and an average period 14~min ($\sigma=5$~min). More detailed analysis of spiking behaviour of oyster fungi \emph{Pleurotus djamor}  has been done in~\cite{adamatzkyspiking}. There  we demonstrated that the fungi generate action potential like impulses of electrical potential, these impulses are often grouped in trains of spikes. The trains can be high-frequency, with average period of a spike of 2.6~min, and low-frequency, with average period of 14~min. There are possible transitions between the high and low frequency spiking.

An electrical response of fruit bodies to short (5~sec) and long (60~sec) thermal stimulation with open flame is analyses in details. We show that non-stimulated fruit bodies of a cluster react to thermal stimulation, with a single action-potential like spike, faster than the stimulate fruit body does.
In a series of scoping experiments, described in \cite{adamatzky2018towards}, we demonstrated that electrical activity recorded on fruit bodies might act as a reliable indicator of the fungi's response to thermal and chemical stimulation.  A stimulation of a fruit is reflected in changes of electrical activity of other fruits of a cluster, i.e. there is distant information transfer between fungal fruit bodies. Thus the fruit bodies can be seen as input port of the fungal computer. The spikes travelling along the mycelium network as units of information. And, logical functions can be implemented via collisions of the travelling spikes at the junction of the mycelium network.  In automaton models of a fungal computer we shown that a structure of Boolean functions realised depends on a geometry of a mycelial network~\cite{adamatzky2018towards}.

The fungal computers implement logical functions via interaction of voltage spikes, travelling along mycelium strands, at the junctions between strands.  Thus, each junction can be seen as an elementary processor of a distributed multi-processor computing network. How many elementary processors could a fungal computer accommodate? Assume that a number of junctions between mycelium strands is proportional to a number of hyphal tips: 10-20 tips per 1.5-3~mm\textsuperscript{3}~\cite{trinci1974study} of a substrate. This means that a cubic meter of the substrate bears up to a billion of elementary processing units. An electrical activity in fungi can be used for communication between distant parts of the mycelium with message propagation speed of 0.5~mm/sec~\cite{olsson1995action}. The speed is much lower than, e.g. an action potential in plants, however this is a critical disadvantage of fungal computers, because their function would be not to compete with fungal computers but complement them by providing an embedded distribution computation in ecological building materials. 

\section{Discussion}

 The ideas proposed in the paper bear at least three substantial risks. First, very few examples of monolithic structures grown with living fungi have been reported in the literature, and risks of collapse, contamination, inhibited growth, inability to control shape all remain under-investigated especially in the context of larger length scale structures than have been demonstrated thus far. Second, we may have a problem with limiting binding capacity to the outer surface or toxicity of the polymers and/or metal nanoparticles. Third, preserving the grown large-scale monolith might be difficult due to environmental conditions. The risks can be counter-argumented by the following findings. 

Structural characteristics of fungal materials are proven, including recent studies on fabrication factors influencing mechanical, moisture-and water-related properties of mycelium-based composites~\cite{appels2019fabrication} to be suitable for construction industry~\cite{girometta2019physico}.  Architectural analysis demonstrated that houses inspired by forest species are feasible, rational and beneficial from the economic/environmental perspectives~\cite{vallas2017using}.

Nano- and micro-particles are intaken by plants~\cite{monica2009nanoparticles,hischemoller2009vivo,koelmel2013investigation} and slime mould~\cite{gizzie2016hybridising,dimonte2015magnetic}, the conductive polymers are intaken by plants~\cite{adamatzky2018computers}. Certain species of fungi bind Fe\textsuperscript{2+}/Cu\textsuperscript{2+}, and possibly other metals as well~\cite{harms2011untapped}. 
An ability to ‘program’ electrical properties of living substrates, on slime mould and plants, by intake carbon nanotubes, graphene oxide, aluminium oxide nanoparticles shown in~\cite{gizzie2016living}. These findings partly mitigate second risk.

With regards to preserving fungal architectures from weather elements we have already developed  unique technology based on partnering with fungi and on employing the mycelium as key ingredient to bind/transform different typologies of substrates, turning them into functional materials~\cite{camere2018fabricating,appels2019fabrication}. We have also demonstrated that innovative architectures co-exist/co-evolve with living substrates~\cite{hamann2017flora}.

\section{Acknowledgement}

This project has received funding from the European Union's Horizon 2020 research and innovation programme FET OPEN ``Challenging current thinking'' under grant agreement No 858132. 

\bibliographystyle{plain}
\bibliography{bibliography}

\end{document}